\documentclass{article}

\usepackage{arxiv}
\usepackage{tabularx}
\newcolumntype{b}{X}
\newcolumntype{s}{>{\hsize=.5\hsize}X}
\usepackage[utf8]{inputenc} % allow utf-8 input
\usepackage[T1]{fontenc}    % use 8-bit T1 fonts
\usepackage{hyperref}       % hyperlinks
\usepackage{url}            % simple URL typesetting
\usepackage{booktabs}       % professional-quality tables
\usepackage{amsfonts}       % blackboard math symbols
\usepackage{nicefrac}       % compact symbols for 1/2, etc.
\usepackage{microtype}      % microtypography
\usepackage{lipsum}
\usepackage{lineno,hyperref}
\usepackage{graphicx, subfigure}
\usepackage{multirow}

\title{Recent advances in deep learning applied to skin cancer detection}

\author{
  Andre G. C. ~Pacheco \\
  Graduate Program in Computer Science, PPGI \\
  Federal University of Espírito Santo, UFES \\
  Av. Fernando Ferrari 514, Vitória-ES, Brazil\\
  \texttt{agcpacheco@inf.ufes.br} \\
   \And
 Renato R. ~Krohling \\
  Production Engineering Department and \\
  Graduate Program in Computer Science, PPGI \\
  Federal University of Espírito Santo, UFES \\
  Av. Fernando Ferrari 514, Vitória-ES, Brazil\\
  \texttt{rkrohling@inf.ufes.br} \\
}

\begin{document}
\maketitle

\begin{abstract}
Skin cancer is a major public health problem around the world. Its early detection is very important to increase patient prognostics. However, the lack of qualified professionals and medical instruments are significant issues in this field. In this context, over the past few years, deep learning models applied to automated skin cancer detection have become a trend. In this paper, we present an overview of the recent advances reported in this field as well as a discussion about the challenges and opportunities for improvement in the current models. In addition, we also present some important aspects regarding the use of these models in smartphones and indicate future directions we believe the field will take.
\end{abstract}

% keywords can be removed

\section{Introduction}
Skin cancer is the most common cancer worldwide. The World Health Organization (WHO) estimates that one in every three cancers diagnosed is a skin cancer \cite{WHO2019}. In countries such as USA, Canada, and Australia, the number of people diagnosed with skin cancer has been increasing at a fairly constant rate over the past decades \cite{CCA2018, CCSsACoC2014, siegel2019cancer}. The deadliest type of skin cancer is the melanoma, and its early detection greatly increases the prognosis of patients \cite{jerant2000early}. Nonetheless, there is a lack of medical instruments and qualified professionals to assist the population, especially in rural areas \cite{feng2018comparison} and in economically emerging countries \cite{scheffler2008forecasting}. Over the past decades, different computer-aided diagnosis (CAD) systems have been proposed to tackle skin cancer detection. These systems are mostly based on traditional computer vision algorithms to extract various features, such as shape, color, and texture, in order to feed a classifier \cite{ercal1994, celebi2007, wighton2011, maglogiannis2015, barata2014}. Recently, machine learning techniques have become a trend to deal with this task. Deep learning models, in particular, Convolutional Neural Networks (CNN), have been achieving remarkable results in this field. Yu et al. \cite{yu2017} presented a very deep CNN and a set of schemes to learn under limited training data. Esteva et al. \cite{esteva2017} used a pre-trained CNN model to train more than 120 thousand images and achieve a dermatologist-level diagnostic. Haenssle et al. \cite{haenssle2018} and Brinker et al. \cite{brinker2019} presented CNN models that have shown competitive or outperformed the dermatologists. Other efforts have been made using deep learning to detect skin cancer, such as ensemble of models \cite{codella2017, harangi2018}, feature aggregation of different models \cite{yu2019}, among others \cite{han2018, attia2017, nida2019melanoma}.

The recent progress achieved by the machine learning methodologies has been leading to the accession of smartphone-based applications as a tool to handle the lack of dermatoscopes\footnote{a medical instrument that allows the visualization of the subsurface structures of the skin revealing lesion details in colors and textures} available to dermatologists and general practitioners. According to the Ericsson mobile report \cite{ericsson2019}, there are around 7.9 billion smartphones around the world. Thereby, a CAD system embedded in smartphones seems to be a low-cost approach to tackle this problem. However, even though this technology has the potential to be widely used in dermatology, there are important aspects that must be addressed such as target users and how to present the system predictions. In addition, there are important ethical concerns regarding patient confidentiality, informed consent, transparency of data ownership, and data privacy protection \cite{chao2017smartphone}.

Since the impact of machine learning in dermatology will increase in the next few years, the goal of this paper is to critically review the latest advances in this field as well as to reflect on the challenges and aspects that need to improve. To this end, first, we present the main methodologies and results reported for the task. Then, we provide a discussion about general limitations regarding machine learning methods and smartphone-based application issues. Lastly, we conclude this paper with our perspectives about this field for the future.

\section{Automated skin cancer detection}
\subsection{Recent advances}
Automated skin cancer detection is a challenging task due to the variability of skin lesions in the dermatology field. The recent advances reported for this task have been showing that deep learning is the most successful machine learning technique addressed to the problem. In this sense, the International Skin Imaging Collaboration (ISIC) has been playing an important role by maintaining the ISIC Archive, an international repository of dermoscopic skin images, which includes skin diseases and skin cancer \cite{isic2019}. This archive has been providing data for different deep learning methodologies such as the ones proposed by Yu et al. \cite{yu2017}, Codella et al. \cite{codella2017}, Haenssle et al. \cite{haenssle2018}, and Brinker et al. \cite{brinker2019}. Currently, the ISIC archive contains 25,331 images for training and 8,238 for testing. This dataset is available for research purposes.

While developing approaches using the ISIC archive is important, it constrains its use for dermoscopic images. It means that this system cannot be used, for example, in smartphone apps, except if the device has a special dermoscope attached to it. In this context, it is necessary to expand the models to also handle clinical images. However, for this case, there is no large public archive available such as ISIC. Thereby, Han et al. \cite{han2018} combined clinical images from 5 repositories, public and private, in order to detect benign and malignant cutaneous tumors. Nonetheless, a breakthrough work was presented by Esteva et al. \cite{esteva2017} in which the authors collected 129,450 clinical images and trained a convolutional neural network (CNN) that achieved a dermatologist level in the benign/malignant identification. Unfortunately, this dataset is private and is not available for the research community.

Another trend in this field is to adopt an ensemble of deep models instead of a single method. The main goal of this approach is to make predictions more effective and reliable. Codella et al. \cite{codella2017} used an ensemble of different deep models, including deep residual networks and convolutional neural networks (CNNs), in order to detect malignant melanomas, the deadliest type of skin cancer. Similarly, Gessert et al. \cite{gessert2018skin} adopted several types of CNN architectures to classify 7 different types of skin diseases. In general, the ensemble of models has been achieving landmark results, particularly for ISIC archive \cite{perez2019solo}. 

In Table \ref{tab:summ}, we summarize all previously mentioned methods and their main contributions. It is important to note that all those models use only images to output their diagnostics. In fact, dermatologists do not trust only on the image screening, they also use the patient demographics in order to provide a more reliable diagnostic. Clinical features such as the patient’s age, sex, ethnicity, if the lesion hurts or itches, among many others, are relevant clues towards a better prediction \cite{wolff2017}. Recently, Pacheco and Krohling \cite{pacheco2019impact} presented a deep model approach that uses images collected from smartphones and patient demographics to detect six different types of skin lesions (three skin diseases and three skin cancers). They achieved an improvement of approximately 7\% by combining both types of data. In alignment with that work, Google Health researchers developed a deep learning system that is able to combine one or more images with the patient metadata in order to classify 26 skin conditions \cite{liu2019deep}. The addition of metadata provided a 4-5\% consistent improvement in their model. They also report a result that is on par with U.S. board-certified dermatologists. Nonetheless, the authors indicate that is necessary to prospectively investigate the clinical impact of using this tool in actual clinical workflows.

\begin{table}
\caption{A summary of the recent deep learning models proposed to skin cancer detection}
\begin{tabularx}{\textwidth}{c|ssb}
\hline
\textbf{Ref.} &
\textbf{Objective} &
\textbf{Model} &
\textbf{Main findings} \\ \hline

%%%%%%%%%%%%%%%%%%%%%%%%%%%%%%%%%%%%%%%%%%%%%%%%%%%%%%%%%%%%%%%%%%%%%%%%%%%%%%%%%%%%%%%%%%%%%%%%%%%%%%%%%%%%%%%%%%%%%%%%%%%%%
\cite{yu2017} & 
Diagnose melanoma and non-melanoma using dermoscopic image &
A two-stage framework composed of a fully convolutional residual network (FCRN) and a Deep Residual Network (DRN) & 
It was one of the first deep learning models applied to skin cancer detection and experimental results demonstrate 
the significant performance gains of the proposed framework compared to handcrafted feature models
\\ \hline
%%%%%%%%%%%%%%%%%%%%%%%%%%%%%%%%%%%%%%%%%%%%%%%%%%%%%%%%%%%%%%%%%%%%%%%%%%%%%%%%%%%%%%%%%%%%%%%%%%%%%%%%%%%%%%%%%%%%%%%%%%%%%

%%%%%%%%%%%%%%%%%%%%%%%%%%%%%%%%%%%%%%%%%%%%%%%%%%%%%%%%%%%%%%%%%%%%%%%%%%%%%%%%%%%%%%%%%%%%%%%%%%%%%%%%%%%%%%%%%%%%%%%%%%%%%
\cite{haenssle2018} &
Diagnose melanomas and nevus using dermoscopic images &
Inception v4 CNN model &
The authors compared the model performance to a group of 58 dermatologists using 100 images in the test set. The model AUC was greater than the average AUC of the dermatologists
\\ \hline
%%%%%%%%%%%%%%%%%%%%%%%%%%%%%%%%%%%%%%%%%%%%%%%%%%%%%%%%%%%%%%%%%%%%%%%%%%%%%%%%%%%%%%%%%%%%%%%%%%%%%%%%%%%%%%%%%%%%%%%%%%%%%

%%%%%%%%%%%%%%%%%%%%%%%%%%%%%%%%%%%%%%%%%%%%%%%%%%%%%%%%%%%%%%%%%%%%%%%%%%%%%%%%%%%%%%%%%%%%%%%%%%%%%%%%%%%%%%%%%%%%%%%%%%%%%
\cite{brinker2018} &
Diagnose melanomas and nevus using dermoscopic images &
ResNet50 CNN model &
The authors compared the model to a group of 157 dermatologists using 100 images. The model outperformed 136 of them in terms of average specificity and sensitivity
\\ \hline
%%%%%%%%%%%%%%%%%%%%%%%%%%%%%%%%%%%%%%%%%%%%%%%%%%%%%%%%%%%%%%%%%%%%%%%%%%%%%%%%%%%%%%%%%%%%%%%%%%%%%%%%%%%%%%%%%%%%%%%%%%%%%

%%%%%%%%%%%%%%%%%%%%%%%%%%%%%%%%%%%%%%%%%%%%%%%%%%%%%%%%%%%%%%%%%%%%%%%%%%%%%%%%%%%%%%%%%%%%%%%%%%%%%%%%%%%%%%%%%%%%%%%%%%%%%
\cite{han2018} &
Diagnose benign and malignant cutaneous tumors among 12 types of skin diseases using clinical images &
ResNet-152 CNN model &
The results achieved by the model were comparable to the performance of 16 dermatologists. The authors also affirm that it is necessary to collect images with a broader range of ages and ethnicities in order to improve the model
\\ \hline
%%%%%%%%%%%%%%%%%%%%%%%%%%%%%%%%%%%%%%%%%%%%%%%%%%%%%%%%%%%%%%%%%%%%%%%%%%%%%%%%%%%%%%%%%%%%%%%%%%%%%%%%%%%%%%%%%%%%%%%%%%%%%

%%%%%%%%%%%%%%%%%%%%%%%%%%%%%%%%%%%%%%%%%%%%%%%%%%%%%%%%%%%%%%%%%%%%%%%%%%%%%%%%%%%%%%%%%%%%%%%%%%%%%%%%%%%%%%%%%%%%%%%%%%%%%
\cite{esteva2017} &
Diagnose 757 types of skin diseases using clinical images &
GoogleNet Inception v3 CNN model &
The model achieved performance on par with 21 dermatologists considering the binary classification of the most common and the deadliest cases of skin cancer
\\ \hline
%%%%%%%%%%%%%%%%%%%%%%%%%%%%%%%%%%%%%%%%%%%%%%%%%%%%%%%%%%%%%%%%%%%%%%%%%%%%%%%%%%%%%%%%%%%%%%%%%%%%%%%%%%%%%%%%%%%%%%%%%%%%%

%%%%%%%%%%%%%%%%%%%%%%%%%%%%%%%%%%%%%%%%%%%%%%%%%%%%%%%%%%%%%%%%%%%%%%%%%%%%%%%%%%%%%%%%%%%%%%%%%%%%%%%%%%%%%%%%%%%%%%%%%%%%%
\cite{codella2017} &
Diagnose melanoma and non-melanoma using dermoscopic images &
An ensemble composed of DRNs, CNNs and Fully CNNs &
The ensemble of models was compared to the average of 8 dermatologists on a subset of 100 testing images and provided higher accuracy and specificity, and an equivalent sensitivity
\\ \hline
%%%%%%%%%%%%%%%%%%%%%%%%%%%%%%%%%%%%%%%%%%%%%%%%%%%%%%%%%%%%%%%%%%%%%%%%%%%%%%%%%%%%%%%%%%%%%%%%%%%%%%%%%%%%%%%%%%%%%%%%%%%%%

%%%%%%%%%%%%%%%%%%%%%%%%%%%%%%%%%%%%%%%%%%%%%%%%%%%%%%%%%%%%%%%%%%%%%%%%%%%%%%%%%%%%%%%%%%%%%%%%%%%%%%%%%%%%%%%%%%%%%%%%%%%%%
\cite{gessert2018skin} &
Diagnose 7 different types of skin diseases using dermoscopic images  &
An ensemble composed of ResNets, Densenets and Senets &
The authors presented a new strategy based on a vast amount of unscaled image crops to generate final predictions. This approach outperforms most of the current models proposed for the ISIC archive 

\\ \hline
%%%%%%%%%%%%%%%%%%%%%%%%%%%%%%%%%%%%%%%%%%%%%%%%%%%%%%%%%%%%%%%%%%%%%%%%%%%%%%%%%%%%%%%%%%%%%%%%%%%%%%%%%%%%%%%%%%%%%%%%%%%%%

\end{tabularx}
\label{tab:summ}
\end{table}

To conclude this section, it is worth noting the recent work developed by Faes et al. \cite{faes2019automated}. In that work, the authors, who do not have any experience with algorithm development, used the Google Cloud AutoML to design several deep learning models for medical images, including skin cancer. They used a partition of the ISIC archive and reported a result comparable to other elementary classification tasks in this section. On the one hand, it is a democratization of deep learning techniques. However, it also raises some questions about ethical principles when using these automated models.

\subsection{Challenges and opportunities}

The models and results summarized in the previous section demonstrate the potential of CAD systems based on deep learning models applied to skin cancer detection. Nonetheless, there are several concerns that must be addressed in order to improve those systems. In this context, the goal of this section is to present a discussion about these concerns as well as indicate challenges and opportunities in this field.

% \subsubsection{Datasets}
\subsubsection{Dataset, bias, and uncertainty}
It is known that to apply deep learning approaches it is necessary a large amount of data. However, collecting medical data, particularly from skin cancer, is a challenging task. Therefore, one of the main concerns of applying deep learning for this task is the lack of training data \cite{han2018, yu2017}. As stated before, the ISIC archive is very important to tackle this issue. However, the number of samples available is still insufficient and very imbalanced among the classes. In order to deal with these problems, several approaches have been proposed, such as transfer learning, data augmentation, up/down-sampling, and weighted loss \cite{vasconcelos2017, perez2018data}. Nevertheless, there is still room for improvement and approaches to learn with limited data and based on weak supervision seem to be good choices to deal with it.

\begin{figure}
    \centering
  \subfigure[Clinical\label{fig:cli}]{
       \includegraphics[width=0.25\linewidth]{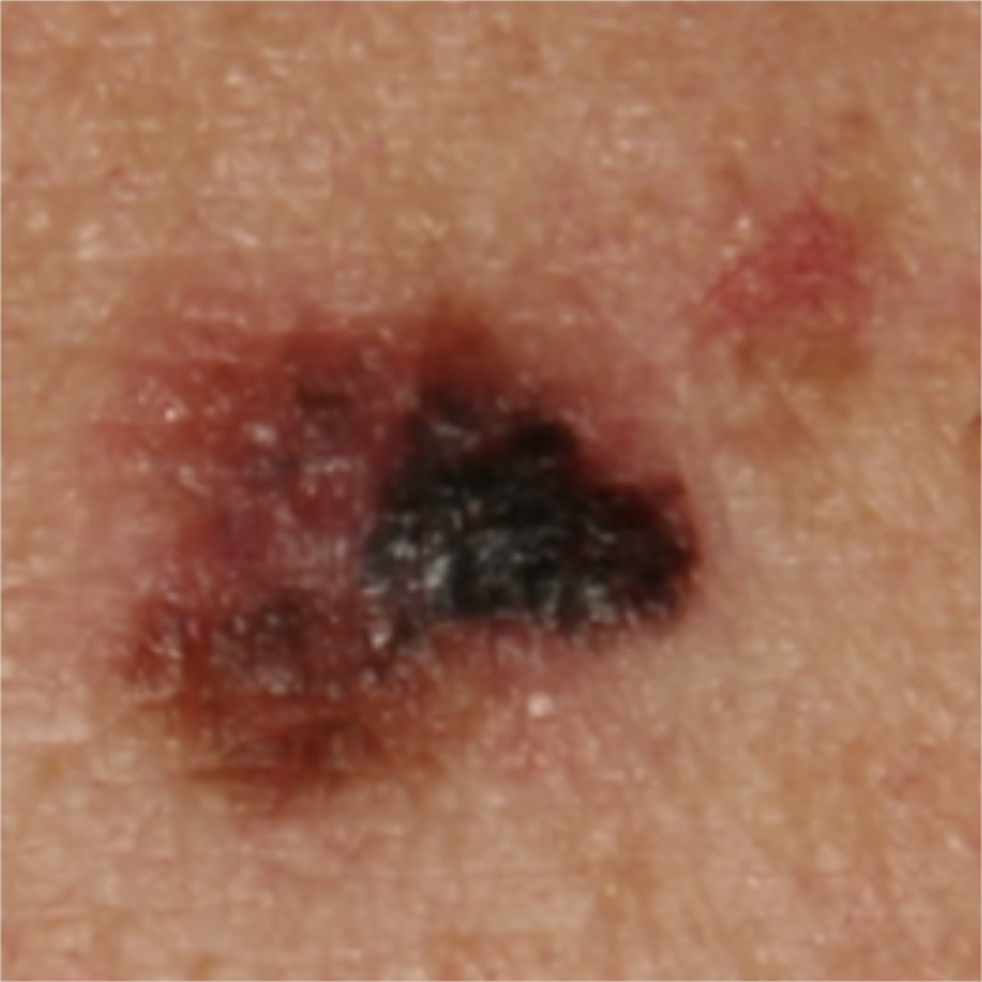}}
    \quad
  \subfigure[Dermoscopic\label{fig:dem}]{%
       \includegraphics[width=0.25\linewidth]{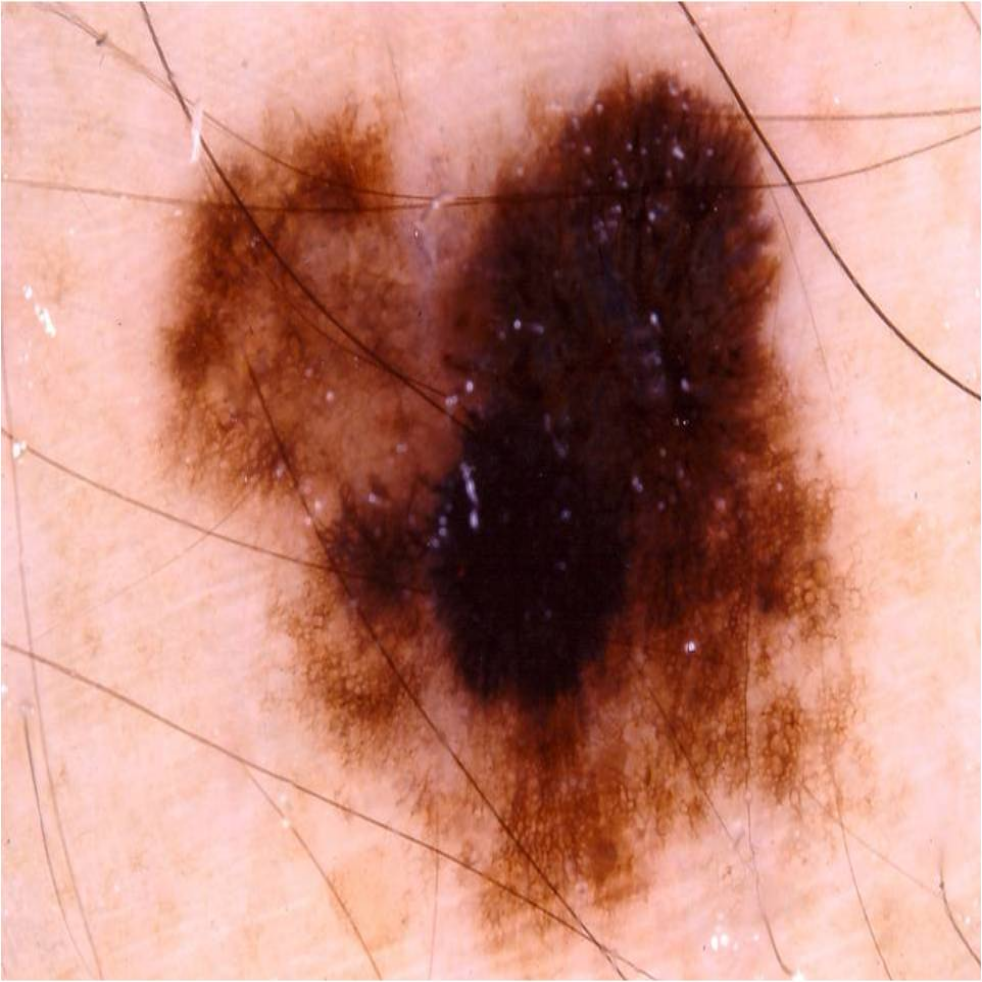}}
    \quad
  \subfigure[Histopathological\label{fig:hist}]{%
        \includegraphics[width=0.25\linewidth]{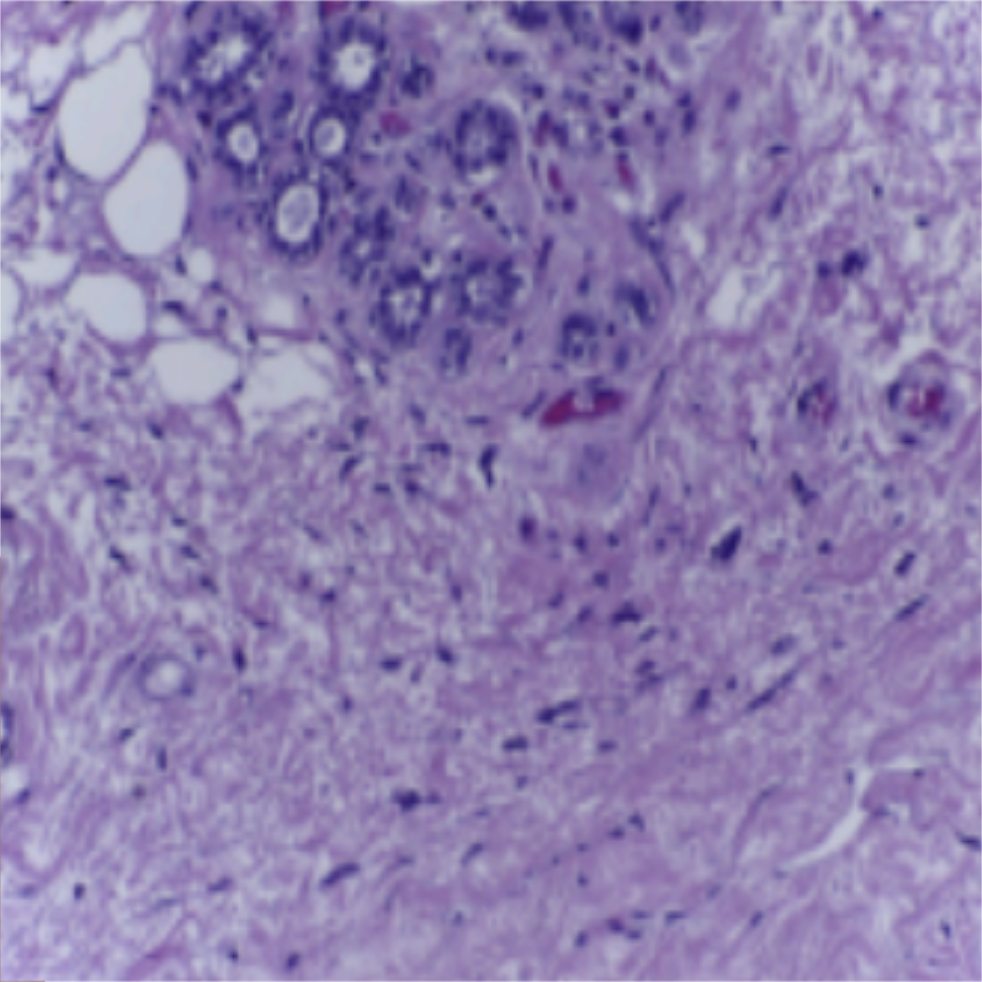}}
        
  \caption{The difference between clinical \cite{han2018}, dermoscopic \cite{isic2019} and histopathological \cite{fondon2018automatic} images of a skin cancer}
  \label{fig:images} 
\end{figure}

It is also important to note that the lack of open clinical data is a limiting factor for this task. As shown in Figure \ref{fig:images}, dermoscopic and clinical images present significant differences related to the level of details available in each image. Thereby, the reuse of a model trained using only dermoscopic images to predict clinical images is not feasible. The previously described works that deal with clinical data either combined some small datasets \cite{han2018} or have access a private ones \cite{esteva2017, liu2019deep}. In this sense, a concerted effort is needed in order to build a clinical image archive such as ISIC. Furthermore, it is important to include, along with the images, the patient demographics (metadata). As Liu et al. \cite{liu2019deep} have shown, the use of metadata may help the deep learning systems deal with the lack of a large number of images.

Another challenge regarding skin cancer detection is to understand the current bias that distorts the performance of the models. Bissoto et al. \cite{bissoto2019constructing} carried out a study that suggests spurious correlations guiding the models. Moreover, some datasets, such as the one used by Liu et al. \cite{liu2019deep}, contain just a few samples of skin types IV and V \cite{wolff2017}, which contribute to the bias. All these points must be considered in order to deploy a model to detect skin cancer for a more diverse group of people.

Beyond the bias, the patient metadata may contain uncertain information. Clinical features such as family cancer history, if the lesion is painful or itching, among many others, are surrounded by uncertainty. Currently, the models do not take it into account, but it is an issue that should be addressed in the future.

\subsubsection{Presenting the predicted diagnosis} \label{sec:present}
Currently, the most common way that models provide the diagnosis is selecting the label that produces the highest probability. Some models also provide a ranking or a threshold for suspicious lesions \cite{han2018, liu2019deep}. However, how can a clinician interpret a low probability assigned to a melanoma? In fact, they require more explanations than only the model’s predictions \cite{zakhem2018should}. Instead of focusing only on the final accuracy, we need to improve how we present the results to the users. In this context, it is very important to determine the target user. Dermatologists, general practitioners, medical students, or even patients, have different levels of knowledge, hence, different needs.

\begin{figure}
    \centering
    \includegraphics[width=0.5\linewidth]{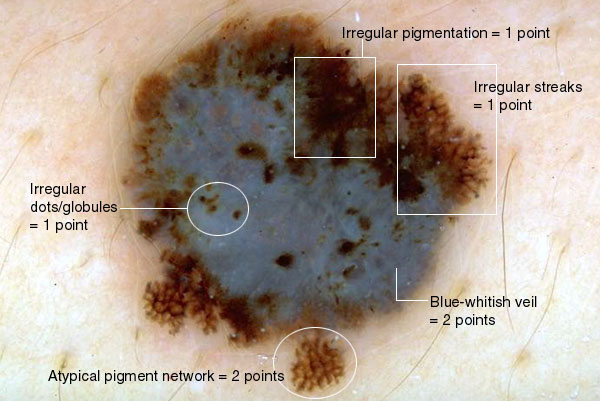}
    \caption{An example of the pattern analysis according to the 7-point checklist \cite{dermoscopy}}
    \label{fig:features} 
\end{figure}

In general, a clinician is interested in CAD systems that support their diagnostic by presenting insights and visual explanations of the features used by the model in the classification process \cite{zakhem2018should}. They want to know why the model is selecting such disease. In this sense, we also need to focus on models that are able to output not only the labels’ probabilities but the pattern analysis as well. Kawahara and Hamarneh \cite{kawahara2018fully} proposed a model to detect dermoscopic feature classification, but it needs to be improved and extended to clinical data. In Figure \ref{fig:features} is depicted an example of the 7-point checklist, an algorithm based on pattern analysis commonly used by dermatologists to detect skin cancer \cite{argenziano1998}. As we can note, the expert is able to identify known patterns in the image in order to determine the final diagnosis. While it is a very challenging task, it should be the ultimate goal of a CAD system employed for skin cancer detection.

\section{Skin cancer detection using smartphones}
As a consequence of the recent progress achieved by CAD systems for skin cancer detection, there are currently several smartphone-based applications that aim to deal with this task. As stated previously, embedding a skin cancer detection in a smartphone is a low-cost approach to tackle the lack of dermatoscopes in remote places. It is clear that this technology has the potential to impact positively on people’s lives. It may accelerate and help clinicians to provide a reliable diagnosis. However, developing such a technology is not only deploying the model in a smartphone. There are important ethical aspects that must be addressed. The amount of those apps available for general users has drawn the attention of different researchers that claim several issues regarding their use. Kassianos et al. \cite{kassianos2015smartphone} carried out a study that identified 40 smartphone apps available to detect or prevent melanoma by non-specialist users. Half of them enabled patients to capture and store images of their skin lesions either for review by a dermatologist or for self-monitoring. Chao et al. \cite{chao2017smartphone} conducted a similar study and concluded that only a few apps have involved the input of dermatologists. In addition, most of them do not provide a disclosure of authorship and credentials. As such, the application should make it clear how it handles user data. It must ensure patient confidentiality as well as let them know what the application does with their data after the model processing. It may sound obvious, but as Chaos et al. \cite{chao2017smartphone} have shown, researchers/developers are not respecting that.

Beyond the problems regarding patient confidentiality and privacy, the lack of regulation for those apps may cause harm to patients or mislead them with an incorrect diagnostic. Let us consider a hypothetical situation of a false negative for melanoma to a given user. It may delay their treatment and, in the worst scenario, it may lead them to death. This is a serious problem that we, machine learning researchers, need to confront. First of all, it is quite important the opinion of dermatologists to improve the effectiveness of this technology. Then, those applications must be exhaustively tested before deployed. Lastly, in our opinion, they should not be allowed to general users before the certification of a board of experts. To this end, it is necessary regulation and we need to advocate for this.

To conclude, in addition to the challenges described in the previous section, in particular, the target users and the way to present the results, there is an important technological issue about deploying deep learning models in smartphones that should be discussed. The main use of this kind of application will be in remote places such as rural areas. In this scenario, it is expected no internet access in those places. However, the current apps do not process the data inside the smartphone, but in a server, which demands internet. There are some fair reasons for this characteristic: the classification is based on more than one model, i.e., an ensemble; the models are computationally expensive, which demands better hardware than the ones usually found in smartphones; and the model’s weights are large files, which may not fit in the smartphone memory. In summary, this is an important aspect that we could not find any discussion about it. In our opinion, this may lead to the development of lighter models in order to deal with it.

\section{Final considerations and future directions}
\begin{figure}
    \centering
    \includegraphics[width=0.70\linewidth]{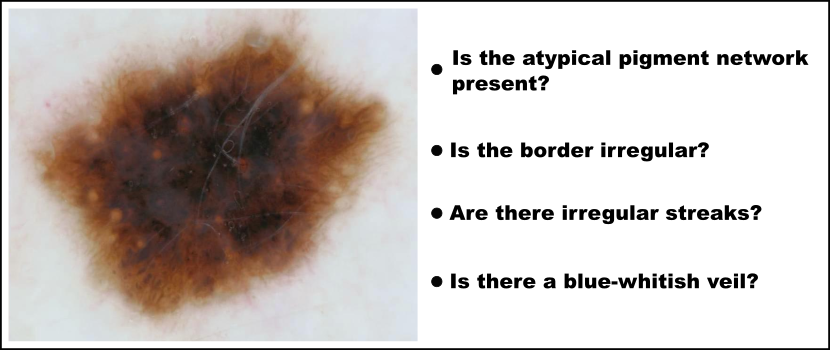}
    \caption{An example of the VQA problem applied to skin cancer detection}
    \label{fig:vqa} 
\end{figure}

Recent advances in deep learning models for skin cancer detection have been showing the potential of this technique to deal with this task. Nonetheless, there are some limitations and important aspects that need to be addressed. In this paper, we presented a discussion about the state-of-the-art approaches as well as the main challenges and opportunities related to this problem. Despite the remarkable results reported, we indicated that there are rooms for improvement, especially for the way the results should be presented. In this context, we believe that in the future this task needs to be addressed as a variant of the visual and question answering (VQA) problem \cite{antol2015vqa}. In Figure \ref{fig:vqa} is illustrated an example of the VQA problem applied to skin cancer detection. The main goal is to allow clinicians to make questions about the lesion in order to understand the predicted diagnosis outputted by the model. This approach is in accordance with the interest of the clinicians, which we described in section \ref{sec:present}. It is clear that addressing skin cancer detection as a VQA problem increases the difficulty of the problem. However, it is an efficient way toward the goal of delivering a more useful tool for doctors.

Another aspect we believe will become a trend in the near future is the use of three types of skin cancer images: clinical, dermoscopic and histopathological. As we can see in Figure \ref{fig:images}, each image presents different characteristics, which may help to correlate features to improve the predicted diagnosis. In addition, CAD systems will be able to act from clinical diagnosis to biopsy, which makes it more desirable and useful. To conclude, regarding the deployment of deep models in smartphones, as noticed earlier, the use of lighter models is necessary in order to make the apps available in remote places. In this context, investigating better ways to improve transfer learning and considering not only the image but also patient demographics are important aspects to be explored in the future.

\bibliographystyle{unsrt}
\bibliography{references}

\end{document}